\documentclass[12pt,preprint]{aastex}
\usepackage{graphicx}
\input epsf

\newcommand{\deltE}{\Delta\kern-1ptE}
\newcommand{\fuse}{{\it FUSE}}

\newcommand{\kms}{km~s$^{-1}$}
\newcommand{\gal}{$\alpha$}
\newcommand{\gb}{$\beta$}
\newcommand{\gla}{$\lambda$}
\newcommand{\etal}{et al.}

\slugcomment{Accepted for publication: ApJ Letters}
\begin{document}

\title{A Hot Wind from the Classical T Tauri Stars:\\ TW Hydrae and T Tauri}

\author{A. K. Dupree\altaffilmark{1,2,3}, N. S. Brickhouse\altaffilmark{1,3}}
\and  
\author{Graeme H. Smith\altaffilmark{4} and Jay Strader\altaffilmark{4}}

\altaffiltext{1}{Harvard-Smithsonian Center for Astrophysics, 
Cambridge MA 02138; adupree@cfa.harvard.edu, nbrickhouse@cfa.harvard.edu}

\altaffiltext{2}{Visiting Astronomer at the Infrared Telescope
Facility, which is operated by the University of Hawaii under
Cooperative Agreement no. NCC5-538 with the National Aeronautics
and Space Administration, Office of Space Science, Planetary Astronomy
Program.} 

\altaffiltext{3}{Guest Investigator, Far Ultraviolet Spectroscopic
Explorer-- a NASA-CNES-CSA FUSE mission operated by the 
Johns Hopkins University; Based in part on 
data from the MAST Archive.}

\altaffiltext{4}{University of California Observatories/Lick Observatory,
Department of Astronomy \& Astrophysics,
University of California, Santa Cruz CA 95064;
graeme@helios.ucsc.edu, strader@ucolick.org}

\begin{abstract}
Spectroscopy of the infrared He I (\gla10830) line with KECK/NIRSPEC
and IRTF/CSHELL 
and of the ultraviolet \ion{C}{3} (\gla 977) and \ion{O}{6} (\gla
1032) emission with \fuse\ 
reveals that the classical T Tauri star TW Hydrae 
exhibits P Cygni profiles, line asymmetries, and absorption   
indicative of a continuous, fast ($\sim$400 km/s), hot
($\sim$300,000 K) accelerating outflow with a mass loss rate
$\sim$10$^{-11}$--10$^{-12}$ $M_{\odot}\ yr^{-1}$ or larger.  
Spectra of T Tauri N appear 
consistent with such a wind.   The source of the emission 
and  outflow seems restricted to the stars themselves. 
Although the mass accretion rate 
is an order of magnitude less for TW Hya than for T Tau, the 
outflow reaches higher velocities at chromospheric temperatures 
in TW Hya. Winds from young stellar objects may be substantially hotter and
faster than previously thought.

\end{abstract}
\keywords{stars:pre-main sequence stars: winds, outflows stars: 
individual (TW Hya, T Tau) infrared: stars ultraviolet: stars}

\section{Introduction}
Young stars with accretion disks 
display energetic jets observed
in optical transitions from low stages of ionization: [O I], [Fe II], 
[S II], [N I], [N II], as well as 
molecular outflows.  The source
of these outflows is 
not well determined.  Shu \etal\ (1994) argue 
they arise from the truncation region of the accretion 
disk where the stellar
magnetosphere, frozen into the disk, causes super-Keplerian rotation and
drives a magnetocentrifugally accelerated wind.  K\"onigl 
\& Pudritz (2000) suggest an extended region of the accretion disk
itself may be responsible.  Or, several 
regions may contribute to a cool wind as indicated by
[Fe~II]: the disk forming a wide angle low velocity component, whereas a high velocity
component is launched from a  region next
to the truncation radius (Pyo \etal\ 2003).  In addition, near UV
and optical lines  show signs of outflow
apparently from the photosphere and low chromosphere (Ardila \etal\ 2002;
Herczeg \etal\ 2002).
To date, the dynamics and energetics
of the outflows in young stars have been principally 
constrained by these  low temperature
species and the modeled winds
are generally cool, T~$<$10$^4$K 
(Shang \etal\  2002).  Beristain \etal\ (2001)
detected broad wings in  He I  5876\AA, and suggested this signaled a 
warmer wind. It is important to determine physical
properties of these winds
and identify their source for they can affect angular momentum
loss, disk structure, and acceleration of optical jets. 

Additional unique spectral features can provide diagnostics of 
the dynamics of outflows from young stars and are reported here.
Two transitions are particularly valuable: the chromospheric
He I (\gla10830; $2s\ ^3S \rightarrow 2p\ ^3P^0$) line 
arising from a metastable state predominantly populated by recombination
following photoionization by the euv continuum.  In a cool
luminous star, this transition is generally formed higher
in the atmosphere (T$\sim$20000K) than H\gal\ and the Ca~II and Mg~II emission
cores (Dupree \etal\ 1992). Because \gla10830 
is not coupled to  local conditions, it is useful to
indicate bulk motions in cool stars including
a handful of young stellar objects (Dupree \etal\ 1992; 
Edwards  \etal\ 2003; Takami \etal\ 2002; Dupree 2004).
Importantly, this \ion{He}{1} line is unaffected by interstellar or
circumstellar absorption.
Another valuable diagnostic 
is the resonance line of \ion{C}{3} (\gla 977), which
has the highest opacity of any of the major ultraviolet
resonance transitions.  
Judged by atomic physics, the opacity of 
\ion{C}{3} \gla977 exceeds all other major far-uv resonance
lines by factors of 3 to 10 (Dupree \etal\ 2005),  
making it sensitive to absorption that could  
reveal the presence of a wind.  In a
collisionally ionized plasma \ion{C}{3}  
signals  temperatures $\sim$ 80,000K (Young \etal\ 2003).
Effects of mass motions causing asymmetries in 
ultraviolet line profiles are widely observed in luminous
cool stars (Dupree \& Brickhouse 1998, Carpenter \etal\ 1999, 
Dupree \etal\ 2005) and the interpretation of
asymmetries in relation to atmospheric dynamics 
is confirmed from semi-empirical modeling
(Lobel \& Dupree 2001).

Two well-studied classical 
T Tauri stars are good targets. 
TW Hydrae and T Tauri have  rotation
axes oriented with 
low inclinations, $<$20$^\circ$, and 
between 8 and 13$^\circ$ respectively (Krist \etal\ 2000;
Herbst \etal\ 1986) so that 
the stellar polar regions  are
observed directly.  The accretion disks are observed face-on.
Although the dipole component of the stellar magnetic field, 
where disk material is thought to be channeled
to the star and thermalized in an accretion shock
(Hartmann 1998 and references therein) may not be aligned with
the rotation axis of the star, it is still in view.  
TW Hya is older (10 Myr), than T Tau, and the accretion rate may be low
[$\sim$4$\times$ 10$^{-10}$ M$_\odot$ yr$^{-1}$, from
H-\gal, (Muzerolle \etal\ 2000) to 10$^{-8}$--10$^{-9}$ M$_\odot$ yr$^{-1}$
from Na D  (Alencar \& Batalha 2002)]
whereas T Tau N, (7.3 Myr)  
has a much higher accretion rate of 3.1--5.7
$\times$10$^{-8}$M$_\odot$ yr$^{-1}$ to 3$\times$10$^{-7}$M$_\odot$ yr$^{-1}$
(Calvet \etal\ 2004; Johns-Krull \etal\ 2000).  Since mass outflows
are thought to be proportional to the mass accretion rate (Calvet 2004), 
different winds might be expected from these
2 stars. 

\section{Observations}

TW Hya was observed at \ion{He}{1}
\gla10830 in July 2002 at KECK II\footnote{
Data  were obtained at the W. M. Keck 
Observatory, which is operated as a scientific partnership among the 
California Institute of Technology, the University of California, and the 
National Aeronautics and Space Administration. The Observatory was made
possible by the generous financial support of the W. M. Keck
Foundation.  The authors recognize and 
acknowledge the very significant cultural role and reverence that the summit 
of Mauna Kea has always had within the indigenous Hawaiian community.  We are 
most fortunate to have the opportunity to conduct observations from this 
mountain.}
using the
NIRSPEC instrument (McLean \etal\ 1998).  Observations were made using
the echelle cross-dispersed mode  with the NIRSPEC-1 order-sorting
filter and a slit of 0.43~$\times$~12 arcsec, yielding a nominal
resolving power of 23,600.  Fringing was minimized by not using 
the two PK50 blocking glass filters.  The exposure time 
totaled 480 seconds. The
spectrum was reduced using the REDSPEC IDL 
software package.  The wavelength scale was established 
using NeArKr arc lamps
and is set to photospheric values using the Si I and 
Mg I absorption as reference.   
He I \gla10830 in T Tau was observed at the IRTF with
CSHELL (Greene \etal\ 1993) in Aug.  1992. 
A 1 arcsec slit was used; the instrumental
resolution is 15 \kms, and the  exposure time was 10
minutes.  This spectrum shows residual fringing near $-$400 \kms.  
Standard IRAF procedures were used to reduce the spectra. 
The wavelength scale, determined by  Ar and Kr lamps, is 
set to the photospheric value, using the
Si I absorption at \gla10844.02. 
Figs.~1 and 2 show that both stars have \ion{He}{1}  P Cygni profiles. 
The absorption extent reaches $-$280 \kms\ in TW Hya, and
$-$220 \kms\ in T Tau.  Since the \ion{He}{1} line is formed at
chromospheric temperatures, these velocities are supersonic
and may be an indication of shocks and transient events.
The photospheric escape velocity is $\sim$500 \kms\  for these stars, 
but at a distance of 1R$_\star$ above the surface, the 
escape speed approaches 300 \kms\ so a
small extension of the atmosphere could easily lead to mass loss.

\fuse\ spectra (Moos \etal\ 2000) 
of TW Hya (Fig. 1), obtained from the MAST archive,  
have a total exposure time of 30.6 ks.  Exposure set C0670101
(15.9 ks) was centered at JD 2452690.834; exposure set C0670102 
(14.7 ks) was taken  one day later. 
We use the total exposure.  
Segments of the SiC2A exposures were examined  to
insure that the channel alignment was in place; segment 1 of C0670102
was discarded because of the short exposure time.  Extractions of 
night-only observations 
were made  using CALFUSE 2.4 to verify that
scattered solar light is absent in the day spectra. Individual 
segments were cross correlated using  stellar
emission lines and summed for  
SiC2A (containing  977\AA) and LiF1A (containing 
1032\AA).  The wavelength offsets are determined using
interstellar absorption features in the spectrum.  For TW Hya,
interstellar absorption of \ion{O}{1}, \ion{C}{2}, \ion{Si}{2} and \ion{Mg}{2} occurs at
a heliocentric velocity of 0$\pm$3 \kms\ (Herczeg et al. 2004), and
the interstellar absorption in the C III \gla977 line has been set
to that velocity with an uncertainty of about 
$\pm$5 \kms.  The radial velocity of
TW Hya is taken as +12.5 \kms (Alencar \& Batalha 2002).  
The wavelength offset for the LiF1A channel was adopted from 
the CALFUSE reduction. Since that channel is used 
for \fuse\ guiding, the offset is the most
reliable; the observed wavelengths of the airglow lines
are equal to their rest values within a few \kms, giving support
to this offset.  The uncertainty in this procedure
is about $\pm$10 \kms.
Far UV spectra of T Tau N were obtained from the \fuse\ archive
(P1630101) where reduction procedures similar to those described
above  were followed. The shortest exposure segments  
were deleted from 14 data segments to achieve 
a total exposure of 19.5 ks.  Interstellar lines observed in
\ion{Mg}{2} at +8 \kms\ (Ardila \etal\ 2002) set the SiC2A
scale by matching to the interstellar \ion{C}{3} absorption.   
The radial velocity of T Tau is taken as +17.5 \kms. 

\section{Evidence for Hot Winds}

The profile of \ion{C}{3} (\gla977) in TW Hya exhibits 
P Cygni structure with  a clear
absorption trough recovering near $-$325 \kms\ and extending 
to higher outflow velocities than the \ion{He}{1} line.  
As expected (Hummer \& Rybicki 1968;
Lobel \& Dupree 2001), a self-absorbed line in a differentially  
expanding atmosphere, appears asymmetric with a steeper
slope occurring on the negative velocity side than on the positive
velocity side of the profile.  A similar shape is found
in the \ion{O}{6} line, although no diminution of absorption creating
an emission `bump' is detected.  Since cool stars lack a   
local continuum in this part of the far uv spectrum, the line 
intensity drops to near zero, exactly 
what is observed in \ion{C}{3} and \ion{O}{6}.  
The similarity of the   
profile shapes in TW Hya suggests that the wind at the \ion{C}{3}
level (80,000K) continues to higher temperatures of 3$\times$10$^5$K 
(Young \etal\ 2003), indicated by the presence of \ion{O}{6} asymmetry, assuming a
collisionally ionized plasma.  The \ion{C}{4} and \ion{N}{5} profiles
of TW Hya (Herczeg \etal\ 2002) also show the same asymmetry as \ion{O}{6}, typical of 
wind absorption.

The presence of wind opacity can be investigated by fitting a Gaussian
profile to the long wavelength wing of the UV  
lines in TW Hya (Fig. 1).  These one-sided  fits predict 
line centers of $-$13$\pm$5  and +3$\pm$9 \kms\ for \ion{C}{3}
and \ion{O}{6} respectively and have similar FWHM 
(\ion{C}{3}: 1.04$\pm$0.02\AA, and \ion{O}{6}: 1.14$\pm$0.04\AA). The 
difference between the observed
profile and the fit reveals the wind absorption.  In Fig. 3, the profiles
are normalized to the local continuum provided by
the Gaussian fit.  The furthest outward extent of the profile reaches
$-$260 \kms\ for \ion{He}{1}, $-325$ \kms\ for \ion{C}{3}, and
$-$440 \kms\ for  \ion{O}{6}.  These 
values are in harmony with the velocity of cooler  material inferred from HST/STIS spectra
to be $-$230 \kms\ from lines of \ion{O}{1}, \ion{C}{2}, and  
\ion{N}{1} (Herczeg \etal\ 2002) suggesting the
accelerating outflow typical of a stellar wind. 
Solar wind models (Hu \etal\ 2000) demonstrate that ions possess different
speeds depending on mass, charge, and wind heating characteristics.

A rough estimate
of the mass loss rate required to produce the absorption profiles
in Fig.~3 
can be derived from the Sobolev optical depth, assuming
$\tau_{Sobolev}$=1 (Hartmann 1998). For an outflow velocity of 
400 \kms\ (the total width of the
\ion{O}{6} absorption), reached over a distance of R$_{star}$, and a
solar oxygen abundance (O/H=8.5$\times$10$^{-4}$) with maximum
fractional ionization for \ion{O}{6} of 0.2, the mass loss rate
follows as: $\dot M_{O\ VI}/\phi$  = 2.3 $\times$ 10$^{-11}$ ({$M_\odot$ yr$^{-1}$})
where $\phi$ is the fraction of the  surface where the 
wind originates.  Similarly for \ion{C}{3}, an acceleration to 300 \kms,
over R$_{star}$ suggests  $\dot M_{C\ III}$/$\phi$ = 1.3 $\times$
10$^{-12}$ ({$M_\odot$ yr$^{-1}$}), taking C/H=3.6$\times$ 10$^{-4}$
and a fractional ionization of 0.8. A wind from high latitude regions
might have $\phi \sim 0.3$. These values are less than the
accretion rate inferred from H-\gal, 
4$\times$10$^{-10}$ {$M_\odot$ yr$^{-1}$} 
(Muzerolle \etal\ 2000). However,
if wind optical depths are much larger, possibly 10$^3$ (Hartmann 1998),
the mass loss rate becomes comparable to the accretion rate.

The far UV spectrum of T Tau is clearly of lesser quality than that
of TW Hya, but  characteristics of the profiles are similar to TW Hya:
namely, a P Cygni \ion{He}{1} line extending to $-$220
\kms, a \ion{C}{3} width less than \ion{O}{6}, and  profiles
consistent with blue asymmetry.    
In T Tau, the far UV lines are  narrower than 
the He I emission which, in a wind model, would
suggest absorption with higher column density than in TW Hya.

\section{Discussion and Conclusions}

It is difficult to determine the source of the UV line emission that
provides the flux for wind scattering, but it must be close to the
star.  Dwarf stars show strong emission from \ion{C}{3}
and \ion{O}{6} (Redfield \etal\ 2002) associated with 
activity, and we expect a similar contribution here.
UV line fluxes generally increase in T Tauri stars which
are undergoing accretion, suggesting an origin in the accretion disk,
X-region, or the shocked accretion column.  The accretion disks in
both sources contain H$_2$, are  dusty, and thus are
unlikely to create and support plasma at $10^5$~K.  The intersection
of the dipole magnetic field from the star and the accretion disk,
perhaps the source of the X-wind, is also unlikely to be responsible 
as it is thought to be ionized and heated by X-rays only to
temperatures of $\sim10^4$K or less (Shang \etal\ 2002). Part of the
UV emission undoubtedly results from the accretion shock.  Profiles 
of emission from highly ionized ions in an accretion shock have not been
reliably calculated; the emission observeed might arise either from
turbulent broadening associated with the shocked, cooling gas, or from
infalling gas from the near-side accretion stream. Non-thermal
broadening could produce intrinsically symmetric profiles as 
modeled here. On the other hand, emission from the
infalling gas on the far-side might be blocked by the star, producing
an intrinsically asymmetric profile. The emission at $-$320 \kms\ in the
\ion{C}{3} line seems to rule out this second case, as we would expect
similar features  from \ion{C}{4}, \ion{N}{5}, and \ion{O}{6}. Furthermore, the
star would preferentially block the highest velocities, not the lowest
velocities, since the accretion stream accelerates toward the stellar 
surface.

T~Tauri stars are frequently associated with optical jets and
Herbig-Haro (HH) objects.  Because our targets are face on, jets along
the line of sight might contribute emission from high temperature material.
However, high excitation HH objects that
contain \ion{C}{4} also exhibit a UV continuum  (Bohm
et al. 1987) that is absent here.  One of the brightest HH objects in
the sky shows neither \ion{C}{3} nor \ion{O}{6} in a {\it Hopkins
Ultraviolet Telescope} spectrum (Raymond \etal\ 1997). FUSE spectra of
HH1 and HH2 obtained from the MAST archive, also give no signs of
these ions. Thus it is not likely that HH objects contribute to the
emission reported here.

The emission components of \ion{C}{3} and \ion{O}{6} are exceptionally
broad.  Both T Tau and TW Hya are slow rotators [v~sin~$i$~=20.9 and
5$\pm$2 \kms, respectively (Hartmann \etal\ 1986; Alencar \& Batalha
2002)], yet the observed emission line widths exceed 
those values (\ion{C}{3}: 102 \kms\ and 202 \kms; \ion{O}{6}: 140
\kms\ and 219 \kms\ respectively).  Accounting for absorption can
nearly double the intrinsic widths. If the emission is attributed to active
regions, the broadening would represent an extreme example
of the weak broadening found in normal stars (Redfield \etal\ 2002). 
Emission associated with
the accretion flow and shock is likely to show turbulent
broadening. We note that the UV line widths are significantly larger
than the X-ray line widths.  If the X-rays from TW Hya are generated
at the accretion shock (Kastner \etal\ 2002), the UV lines may
not be directly associated with the shock. On the other hand,  
studies of X-ray emission in young star clusters, suggest that the
strength of the X-ray emission is correlated with stellar rotation, thus
casting doubt on an accretion origin for the X-rays (Stassun \etal\
2004).
 
Whatever the source of these collisionally excited photons, originating on 
or close to the star, they appear to be  scattered  in the outflowing
plasma, producing a sequence of similar P Cygni profiles ranging
from \ion{He}{1} and \ion{C}{3} to \ion{O}{6}.  In our interpretation,
the wind signatures indicate
a continuous outward acceleration  from approximately
the photospheric velocity to several hundred \kms, and
reach  temperatures of 3 $\times$10$^5$ K.  A  
cool  wind has been suggested in TW Hya 
(Herczeg \etal\ 2002, 2004) as a component in the complex Lyman-$\alpha$
profile, and also from the weakness of  one H$_2$  line possibly 
affected by C II absorption. 
Other H$_2$ transitions subject to wind absorption
may offer insight into the configuration of the
stellar wind if the site  of the fluoresced H$_2$ can be
identified. A high temperature, fast  wind might contribute to the opacity
needed (Stassun \etal\ 2004) for the absorption of X-rays in accreting
systems. In addition, a fast hot wind may influence the diminution
of dust in accretion disks  (Alexander \etal\ 2005).

Detailed semi-empirical modeling will be necessary
to derive meaningful mass loss rates.    The velocities
observed in TW Hya are larger than in T Tau, although the wind 
may be less opaque,  which is surprising
if wind characteristics are related to the  accretion rate.  
 
The profiles reported here are
consistent with 
the presence of hot winds reaching escape velocities, that 
may lead to shocks and  optical jets as the wind decelerates  
at greater distances from the star.   The polar orientation of
TW Hya and T Tau enabled full access to regions likely to
contain open field lines enhancing mass outflow. 
Observations of disk-star systems at 
various inclination angles can help to define  the
structure of both the emission region and the wind. 

We wish to thank Ian McLean and the NIRSPEC team for both the development of 
this instrument and the REDSPEC reduction software. We also thank Lee
Hartmann for useful discussions. J.S. acknowledges support from an NSF
Graduate Research Fellowship.

%FIGURE 1
\begin{figure}
\begin{center}
\includegraphics[scale=0.6]{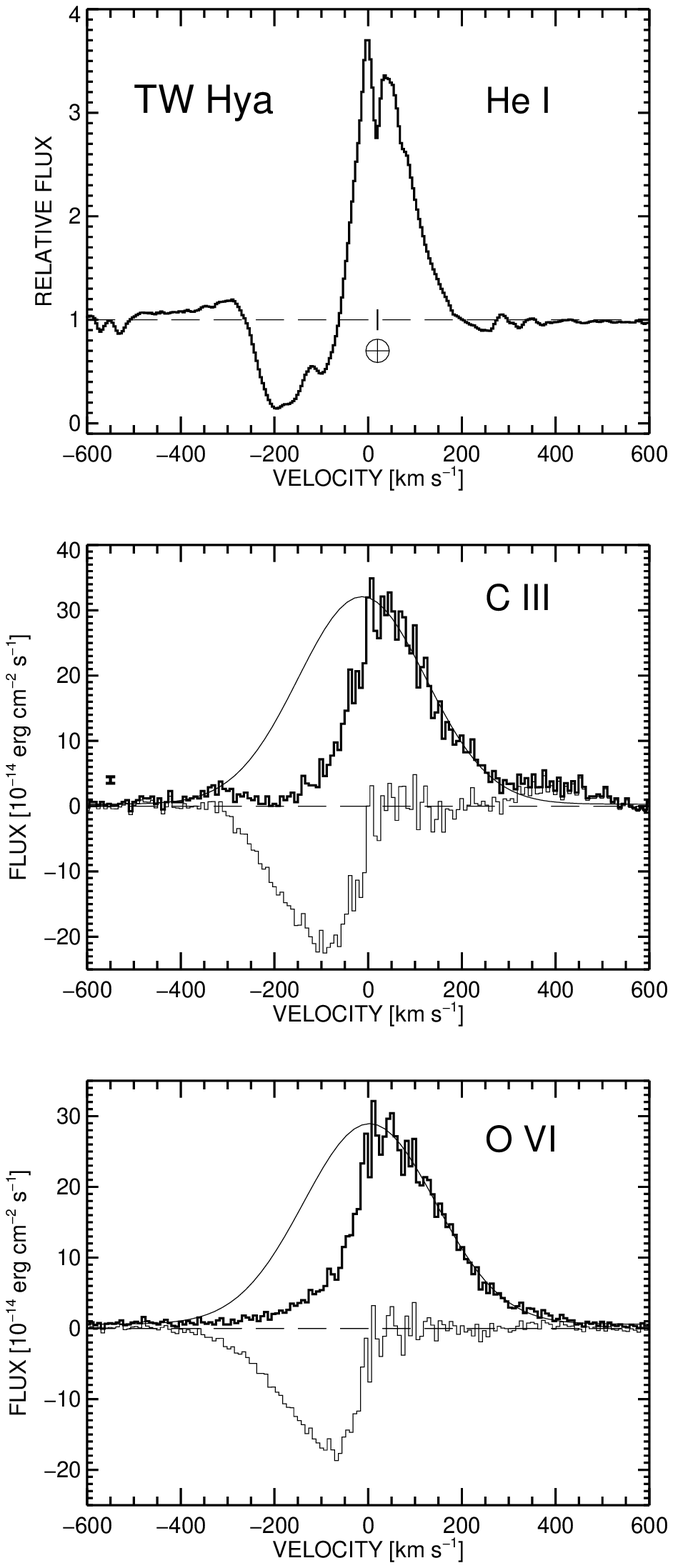}
\caption{\ion{He}{1} \gla10830, \ion{C}{3} \gla977, and
\ion{O}{6} \gla1032 transitions in TW Hya.  The notch in the TW Hya spectrum of \ion{He}{1} 
at $-$100 \kms\  is found also
at times in H-\gal\ profiles (Alencar \& Batalha 2002).
The ultraviolet line
profiles have been fit on the positive velocity side by
a gaussian profile. The difference between the observed profile
and the gaussian fit is shown below the original spectrum. Extraction of the spectra taken during
the \fuse\ night, demonstrates that
the low level emission near \ion{C}{3} centered at $-$320 \kms\ and +400 \kms\ is
associated with the star, and is not contaminated by airglow
emission. The emission feature at $-$320 \kms\ in the \ion{C}{3} profile 
has a 6$\sigma$ significance in one bin sampled twice per resolution 
element. Emission present at 
+400 \kms\ in the \ion{C}{3} profile, might originate from the
star itself (although a full extent of the line
to $\pm$500 \kms\ may be excessive) or it could be due to \ion{O}{1} (\gla978.624) that is 
fluoresced by the stellar \ion{C}{3} line, and arising in the
expanding wind with an outflow of $\sim$100 \kms. A similar feature
is identified in the \fuse\ spectrum of the cool supergiant
\gb\ Dra (Dupree \etal\ 2005).
 }
\end{center}
\end{figure}

%FIG 2

\begin{figure}
\begin{center}
\includegraphics[scale=0.6]{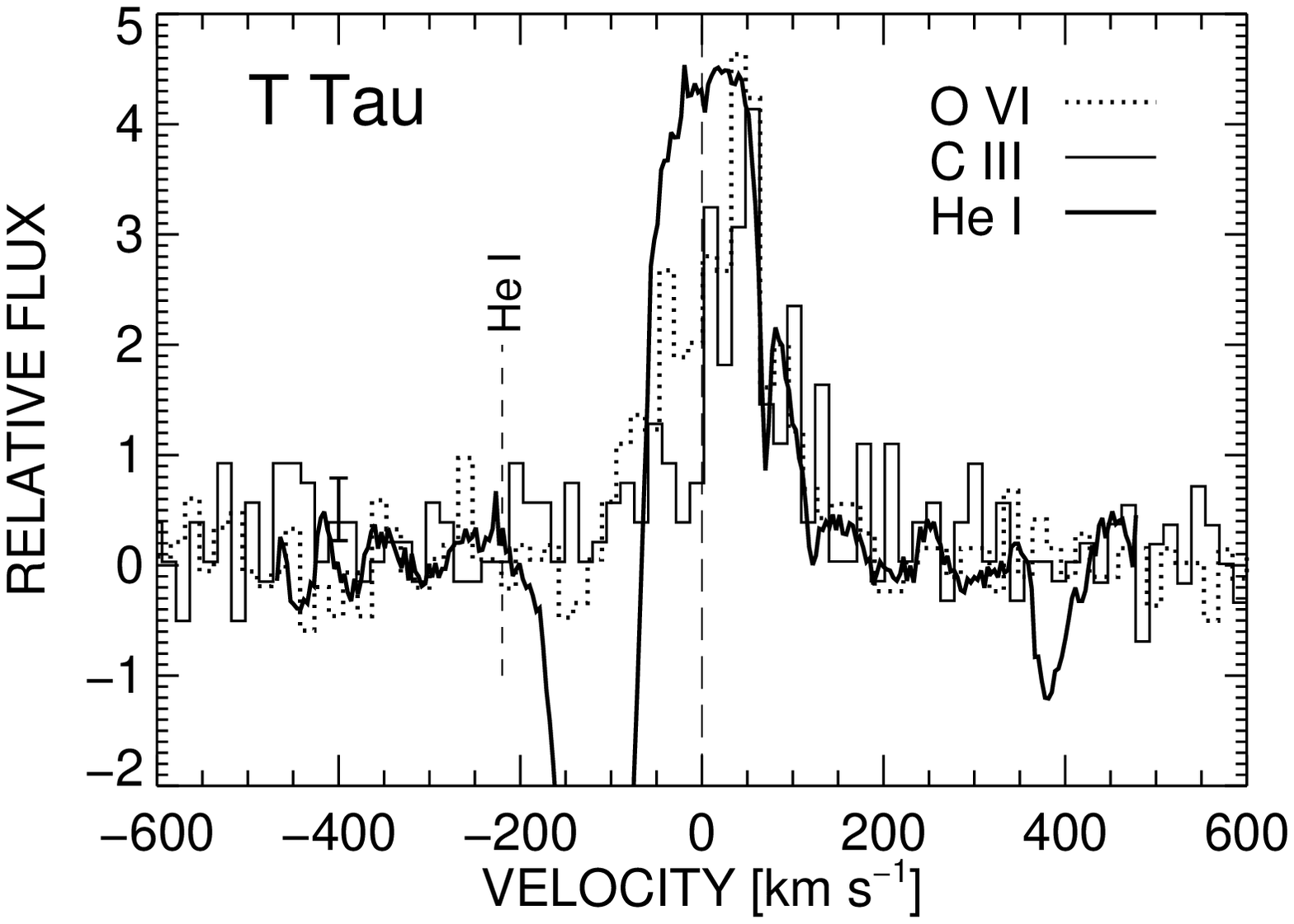}
\caption{\ion{He}{1}, \ion{C}{3}, and \ion{O}{6} in T Tauri. FUSE spectra are binned
to a resolution element.  The short dashed line
marks the maximum outflow velocity in the He I line at $-$220 \kms.  
Fluoresced H$_2$ 
emission surrounds T Tau (Walter \etal\ 2003),
and there may be narrow absorption in the \ion{C}{3} profile,
$-$144 \kms\ and $+$370 \kms\ if at the radial velocity of the star.  
The presence of weak airglow emission in the region between
$-$100 and $-$200 \kms\ in the \ion{C}{3} profile can not be excluded. }
\end{center}
\end{figure}
%FIG3

\begin{figure}
\begin{center}
\includegraphics[scale=0.6]{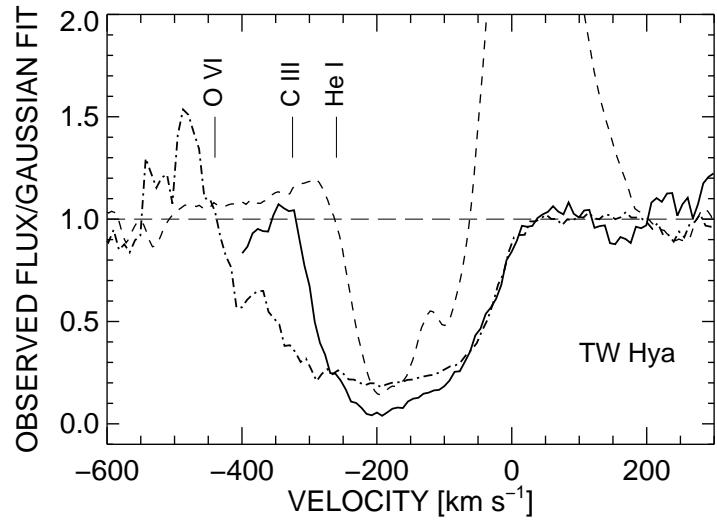}
\caption{Absorption  apparent in \ion{He}{1} \gla10830 ({\it
dashed curve}), \ion{C}{3} \gla977 ({\it solid curve}),
and \ion{O}{6} \gla1032 ({\it dash-dot curve}) lines in TW Hya.  The \ion{C}{3} and
\ion{O}{6} 
absorption features are constructed by using the gaussian
fit (Fig. 1) as the local continuum and calculating the 
ratio: Observed Flux/Fit.  
The \ion{He}{1} line is simply  the observed profile. 
The maximum  velocity extents of  \ion{O}{6}, \ion{C}{3},
and \ion{He}{1}  are marked by vertical lines.
}
\end{center}
\end{figure}

\end{document}